# An inexpensive system for the deterministic transfer of 2D materials

Qinghua Zhao[1,2,3], Tao Wang[2,3], Yu Kyoung Ryu[1], Riccardo Frisenda[1] and Andres Castellanos-Gomez[1,*]

[1] Materials Science Factory. Instituto de Ciencia de Materiales de Madrid (ICMM-CSIC), Madrid, Spain.
[2] State Key Laboratory of Solidification Processing, Northwestern Polytechnical University, Xi'an, P. R. China.
[3] Key Laboratory of Radiation Detection Materials and Devices, Ministry of Industry and Information Technology, Xi'an, P. R. China

andres.castellanos@csic.es

The development of systems for the deterministic transfer of two-dimensional (2D) materials have undoubtedly contributed to a great advance in the 2D materials research. In fact, they have made it possible to fabricate van der Waals heterostructures and 2D materials-based devices with complex architectures. Nonetheless, as far as we know, the amount of papers in the literature providing enough details to reproduce these systems by other research groups is very scarce in the literature. Moreover, these systems typically require the use of expensive optical and mechanical components hampering their applicability in research groups with low budget. Here we demonstrate how a deterministic placement system for 2D materials set up with full capabilities can be implemented under 900 € which can be easily implemented in low budget labs and educational labs.

Key words: 2D materials, deterministic transfer, heterostructures fabrication

After the isolation of graphene and other two-dimensional (2D) materials in 2004-2005,[1,2] the development of the transfer methods to deterministically place 2D materials at specific locations with high accuracy is one of the most important breakthrough in the 2D materials research.[3–8] In fact, these deterministic transfer methods have made possible the fabrication of artificial materials by the stacking of dissimilar 2D materials on top of each other in what is called nowadays van der Waals heterostructures.[9–15]

The implementation of experimental setups for the deterministic placement of 2D materials, however, typically requires costly optical and mechanical components, hampering their implementation in labs with modest budget and their use in science education and public demonstrations. Although, some time ago some of the authors of this work reported all the details to build up a transfer setup with an approximate cost of 7000-8000 € (much cheaper than commercially available systems or conventional transfer setups based on retrofitted metallurgical microscopes or probe stations)[8] this cost can be still a big handicap for the applicability of the system. Here we report all the details to install a fully functional deterministic transfer setup at a total cost under 900 € and with a very compact footprint. The performance of this system is illustrated by placing a few-layer thin InSe flake bridging two electrodes.

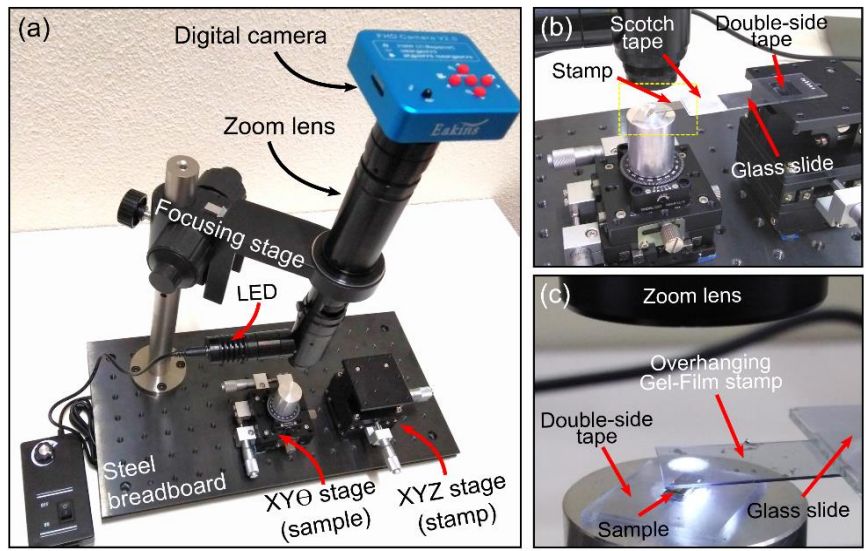

**Figure 1.** (a) Picture of the assembled low-cost system to transfer 2D materials, highlighting some of the key components and details used for deterministic transfer process. (b) Zoomed in picture showing details of the stamp clamping and fixture to the XYZ stage. (c) Zoomed in image of the area highlighted with a dashed yellow rectangle in (b) where the sample and stamp fixture are displayed.

**Figure 1(a)** shows a picture of the presented system to transfer 2D materials. The system is basically composed of a zoom lens with coaxial illumination, a XY+rotation manual stage (sample/substrate stage) and a XYZ manual stage (stamp stage). The zoom lens is supplemented with a 21-megapixel digital camera with HDMI output and all the components are mounted on a magnetic breadboard. The manual stages are attached to the breadboard through magnets glued at the base of the stages. Figure 1(b) shows a close-up picture of the sample and stamp stages where it is shown how the stamp is mounted. We employ a rectangular piece (5 mm by 10 mm approx.) of Gel-Film (WF x4 6.0 mil, by Gel-Pak®) as viscoelastic stamp. Unlike in our previous work, in where we used the PF Gel-Film, we now use the WF Gel-Film that has the polydimethylsiloxane (PDMS) - based gel material bonded to a flexible and quasi transparent backing polyester substrate. The Gel-Film stamp is glued to a glass slide with Scotch tape, leaving most of the stamp overhanging (protruding from the glass slide as shown in Figure 2(c)). We have found that this 'cantilever-like' geometry of the stamp is more advantageous for the deterministic placement of 2D materials than the stamp geometry used in our early work in Ref. [[8]]. Then double-side tape (Scotch® Restickable Tabs) is used to fix the glass slide to the stamp stage and the sample to the sample stage.

**Table 1** summarizes all the different parts needed to implement the system and the reader will find a thorough step-by-step guide to assemble it in the Supporting Information. A list of consumables required to use the deterministic transfer system is shown in **Table 2**.

**Table 1.** Summary of the components of the proposed system. (Note: hyperlinks for the different items are available in the online version of the manuscript).

| | Description | Distributor | | Price (€) |
|---|---|---|---|---|
| **System base** | Ferromagnetic steel optical breadboard | Standa | 1BS-2040-015 | 130.00 |
| | Rubber damping feets (set of 4) | Thorlabs | RDF1 | 4.92 |
| **Imaging system** | Ø25.0 mm pillar post | Thorlabs | RS300/M | 53.25 |
| | Mounting post base | Thorlabs | PB1 | 22.44 |
| | Focusing stage for zoom lens | Aliexpress | | 56.96 |
| | 400× zoom lens with coaxial illuminator | Aliexpress | | 174.83 |
| | Auxiliary 3.5x magnification lens | Aliexpress | | 34.36 |
| | 21 MPix digital camera | Aliexpress | | 74.70 |
| | 22" HDMI monitor | Amazon | | 89.99 |
| **Sample and stamp stages** | Manual rotation stage (for sample) | Thorlabs | MSRP01/M | 66.53 |
| | XY manual stage (for sample) | Banggood | | 63.14 |
| | XYZ manual stage (for stamp) | Banggood | | 81.18 |
| | Magnets (for sample and stamp stages) | Amazon | | 7.15 |
| | | | | **859.45** |

**Table 2.** Summary of the consumables (optional ones marked with *). (Note: hyperlinks for the different items are available in the online version of the manuscript).

| Description | Distributor | | Price (€) |
|---|---|---|---|
| Gel-Film sheet (viscoelastic stamp) | Gel-Pak | WF-30-x4-6mil | 5.00 |
| Nitto tape (for exfoliation of layered crystals) | Amazon | SPV224 | 15.00 |
| Scotch restickable tabs (to attach sample and stamp to their stages) | Amazon | | 5.11 |
| Cyanocrylate based glue | Amazon | | 4.35 |
| Pre-patterned electrodes (20x) * | Ossila | S403A2 | 60.00 |
| Natural bulk 2D crystals (graphite, MoS$_2$, mica…) * | Mineral shop | | 1-100 each |
| Artificial bulk 2D crystals * | HQ Graphene, 2D Semiconductors, 2D-material | | 200-600 each |

In order to illustrate the potential of this inexpensive transfer setup, **Figure 2** shows how one can use this deterministic transfer setup to fabricate a 2D material based field effect transistor. The process starts by exfoliating a bulk crystal of a layered material (in this case a synthetic bulk crystal of InSe, an n-type semiconductor) with Nitto SPV224 tape. The crystals adhered to the surface of the tape are then transferred to the surface of the Gel-Film stamp by gently pressing the tape against the Gel-Film surface and suddenly peeling off the tape. Then the surface of the Gel-Film stamp is inspected under an optical microscope to find the flake that the operator wants to transfer. Note that the stamp can be inspected under the zoom lens imaging system of the transfer setup to identify the flakes if a metallurgical microscope is not available. See the Supporting Information for a comparison between the image quality obtained with the zoom lens system of the transfer setup and with a metallurgical microscope.

As a target substrate we use a SiO$_2$/Si substrate (285 nm of SiO$_2$) with pre-patterned electrodes (Ossila®) separated by 30 µm (step 1). The InSe flake that we want to transfer can be found under the inspection of the zoom lens imaging system by focusing on the surface of the stamp (step 2). The stamp is brought closer to the surface of the sample and the flake is aligned with the electrodes employing the XYZ stamp stage micro-positioner (step 3). The stamp is gently lowered until it contacts the surface of the sample (step 4). Then the stamp is slowly peeled off (steps 5 to 7) until it is completely removed and the InSe has been successfully transferred bridging the two electrodes (step 8). The peeling off process is taken place by lifting up the Z-micromanipulator of the stamp stage at an approximate speed of 0.25 – 0.50 mm/min. Note that these values are a rough indication as the precise lifting speed is usually adjusted by inspecting the peeling off process with the imaging system during the lifting.

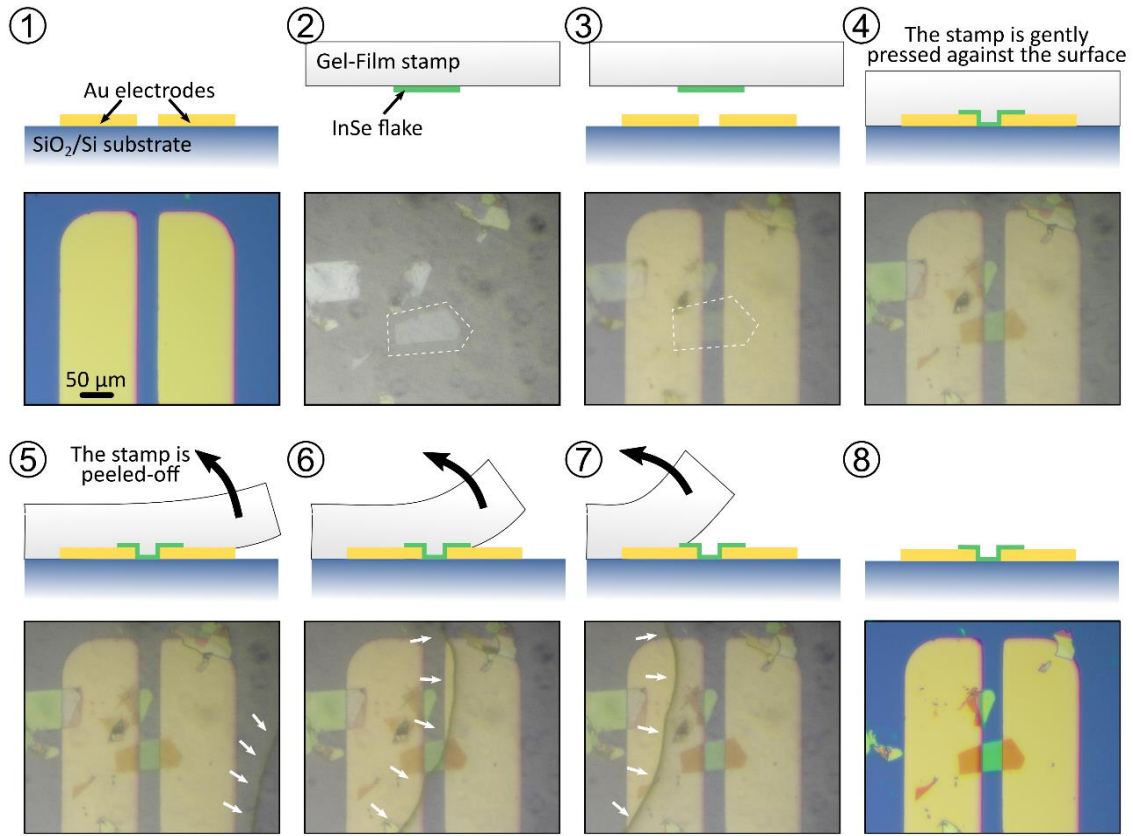

**Figure 2.** Schematics and transfer process of bridging a thin InSe flake onto the pre-patterned gold electrodes on a 285 nm SiO$_2$/Si substrate, including empty pre-patterned gold electrodes (1), isolated thin InSe flake on the surface of the Gel-Film stamp (2), alignment of the flake with the electrodes (3), the stamp is gently pressed against the target sample (4), the stamp is slowly peeled off (5, 6, 7, white arrows indicate the stamp/air interface) and final device (8).

Another widespread use of systems for the deterministic placement of 2D materials is the fabrication of van der Waals heterostructures. In **Figure 3** we demonstrate that our inexpensive deterministic transfer system can be also used to fabricate van der Waals heterostructures by fabricating a fully encapsulated InSe flake between two hexagonal boron nitride (h-BN) flakes. Figure 3a shows the sequence of transfer. First, a h-BN flake is transferred in the middle of a pre-pattern cross-hair marker structure (left column). Second, a InSe flake is transferred onto the bottom h-BN flake (middle column). Finally, another h-BN flake is transferred sandwiching the InSe flake between h-BN sheets (right column in Figure 3a). Figure 3b shows the resulting van der Waals heterostructure.

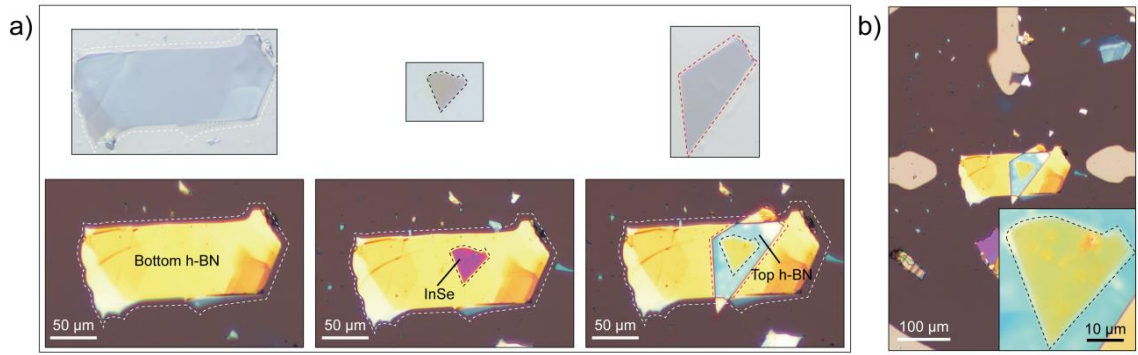

Figure 3. Fabrication process of a fully encapsulated InSe flake heterostructure with h-BN onto a pre-defined location on a SiO$_2$/Si substrate. (a) Sequence of stacking bottom h-BN (left), thin InSe (middle) and top h-BN (right) by the deterministic transfer method. (b) The overall resulting h-BN/InSe/h-BN heterostructure in the middle of a pre-patterned cross-hair and a zoom-in image of the fully encapsulated InSe flake (inset panel).

Finally, we demonstrate how one can use this inexpensive transfer setup to fabricate not only simple devices nor isolated van der Waals heterostructures but also fully-encapsulated electronic devices. Hexagonal boron nitride (h-BN) is an atomically flat insulator with a very low density of Coulomb scattering centers and thus by sandwiching a 2D material between two layers of h-BN provides a very clean dielectric environment preserving the intrinsic electronic properties of the encapsulated material. Because of this, the best performances have been reported for electronic devices based on 2D materials encapsulated between h-BN layers.[16,17] **Figure 4** shows the steps needed for the assembly of a fully encapsulated InSe device using our transfer setup. A bottom narrow h-BN flake is transferred bridging two pre-patterned gold contacts, then a long InSe flake is transferred on top of the h-BN flake protruding and contacting the two gold contacts. Finally, a large h-BN flake is transferred on top. The result is a gold contacted InSe device, fully encapsulated between h-BN flakes.

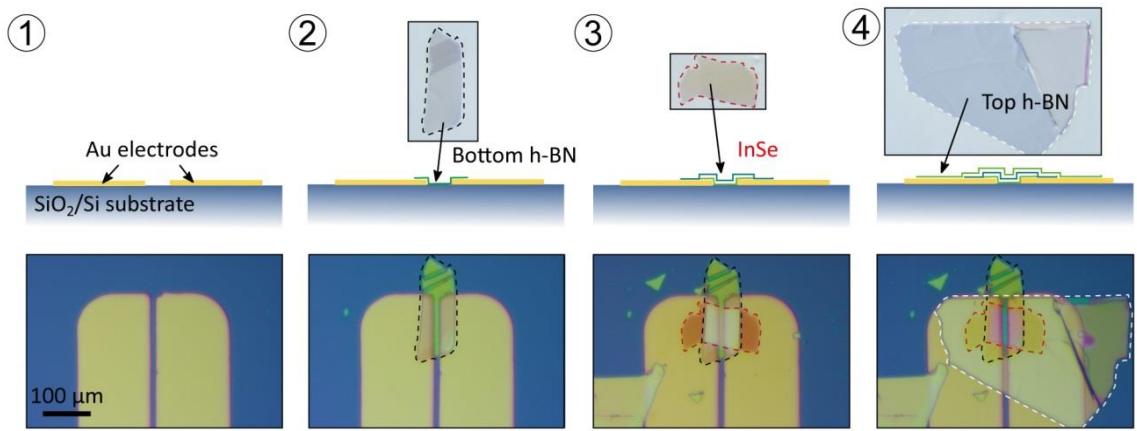

**Figure 4.** Fabrication steps of a InSe device fully encapsulated between h-BN flakes. Pre-patterned gold electrodes on SiO$_2$/Si substrate (1), bottom h-BN flake is transferred (2), InSe flake is transferred bridging the two gold electrodes (3) and transfer of a top h-BN to complete the full encapsulation.

**Conclusions**

In summary, we presented an inexpensive system to transfer 2D materials that can be easily implemented in labs with low budget and educational labs and it can be used for public demonstrations. Moreover, we believe that this transfer system can also help to reduce the entry threshold to work in the field of van der Waals heterostructures. The whole system can be assembled for less than 900 € and the final system has a very compact footprint (easy to transport for educators and public demonstrations). We demonstrate that despite the low cost of the system, it has a functionality comparable to that of more expensive setups. In fact, we show how the system can be used to fabricate devices based on 2D materials and it could even allow a motivated physics teacher to build basic, yet functional devices like transistors or solar cells out of two-dimensional crystals, together with his class.

**Acknowledgements**

We would like to thank the input received from the Reviewers that have improved substantially this manuscript. This project has received funding from the European Research Council (ERC) under the European Union's Horizon 2020 research and innovation programme (grant agreement n° 755655, ERC-StG 2017 project 2D-TOPSENSE). R.F. acknowledges the support from the Spanish Ministry of Economy, Industry and Competitiveness through a Juan de la Cierva-formación fellowship 2017 FJCI-2017-32919. QHZ acknowledges the grant from China Scholarship Council (CSC) under No. 201700290035.

# Supporting Information:

# An inexpensive system for the deterministic transfer of 2D materials

Qinghua Zhao[1, 2, 3], Tao Wang[2, 3], Yu Kyoung Ryu[1], Riccardo Frisenda[1] and Andres Castellanos-Gomez[1, *]

[1] Materials Science Factory. Instituto de Ciencia de Materiales de Madrid (ICMM-CSIC), Madrid, Spain.
[2] State Key Laboratory of Solidification Processing, Northwestern Polytechnical University, Xi'an, P. R. China.
[3] Key Laboratory of Radiation Detection Materials and Devices, Ministry of Industry and Information Technology, Xi'an, P. R. China

andres.castellanos@csic.es

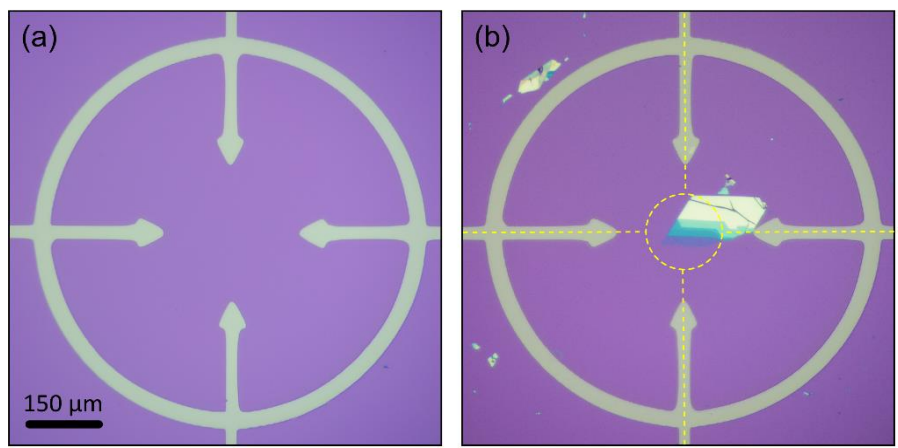

**Figure S1.** Optical images acquired with the transfer setup imaging system before (a) and after (b) transferring an ultrathin $MoS_2$ flake in the middle of the pre-patterned crosshair.

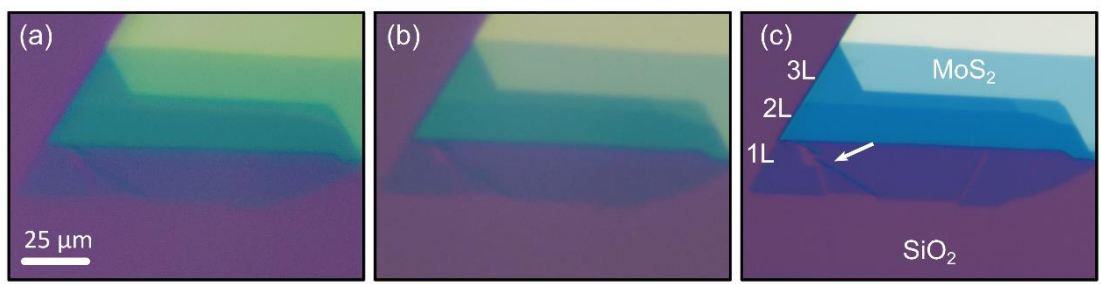

**Figure S2.** Comparison of optical images of the same $MoS_2$ flake shown in Figure S1, acquired with different imaging systems: (a) zoom-lens and camera used in this work (~500 €, WD = 20 mm), (b) Optem 7x zoom lens with 3x mini-tube and Canon EOS 1200D camera (~3100 €, WD = 32 mm) and (c) Motic BA 310 MET-T metallurgical microscope with AMScope MU18 camera (~4500 €, WD = 11 mm). Note that features smaller than 2-3 microns cannot be resolved with our low-cost imaging system and can be barely resolved with more expensive zoom lens systems (see the cracks and fold in the single-layer part of the flake indicated with a white arrow in (c)).

# Step-by-step assembly guide

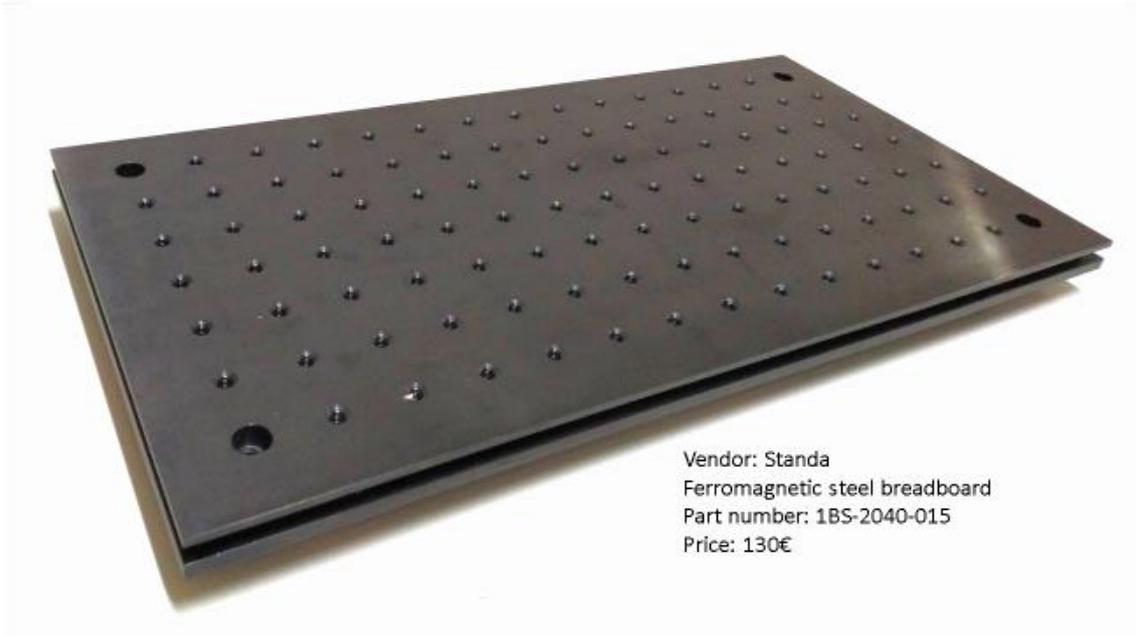

Vendor: Standa
Ferromagnetic steel breadboard
Part number: 1BS-2040-015
Price: 130€

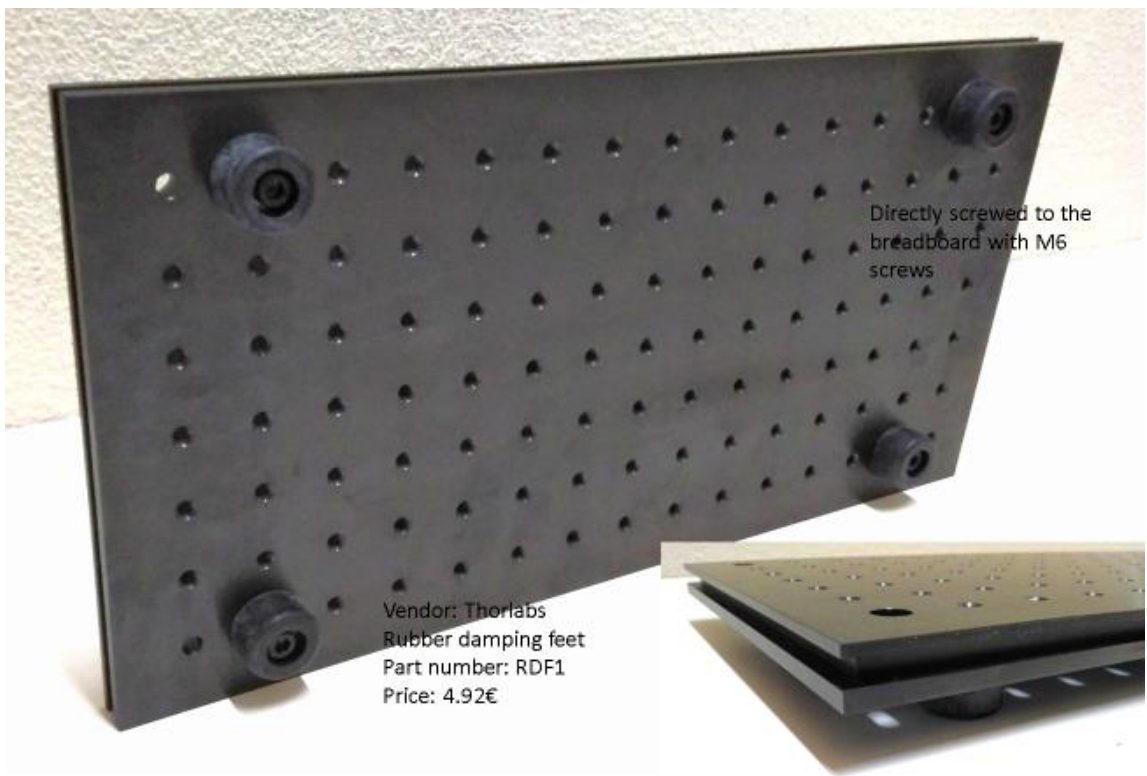

Directly screwed to the breadboard with M6 screws

Vendor: Thorlabs
Rubber damping feet
Part number: RDF1
Price: 4.92€

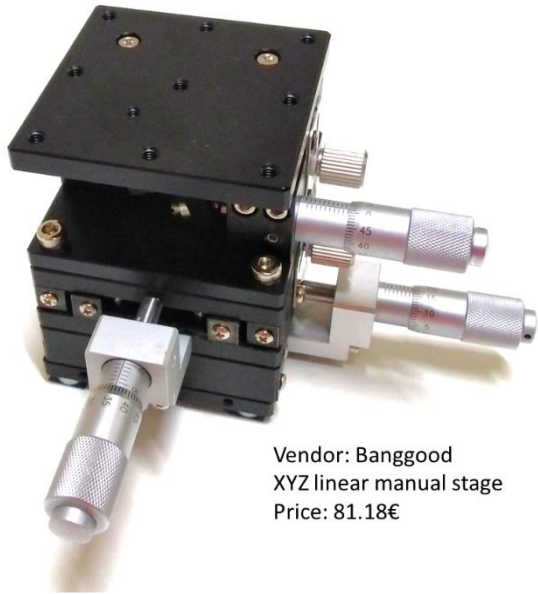

Vendor: Banggood
XYZ linear manual stage
Price: 81.18€

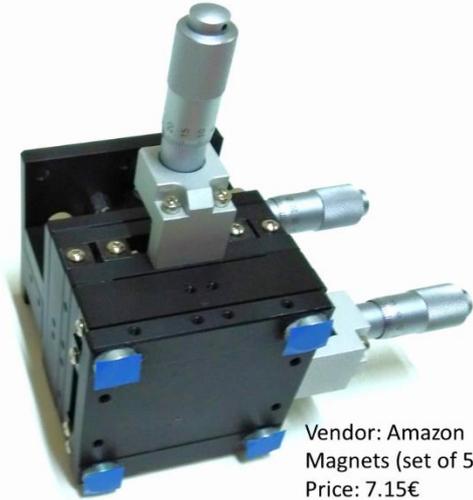

Vendor: Amazon
Magnets (set of 50)
Price: 7.15€

4 magnets are glued on the base of the manual stage (with epoxy glue, Nural 26) and a layer of insulator tape is attached.

The manual rotation stage is glues (cyanoacrylate) onto the XY stage.

Vendor: Thorlabs
Manual rotation stage
Part number: RMSRP01/M
Price: 66.53€

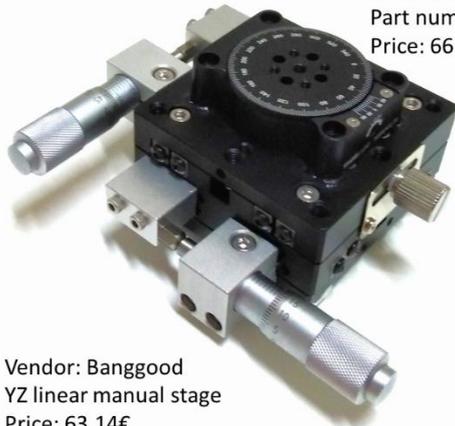

Vendor: Banggood
YZ linear manual stage
Price: 63.14€

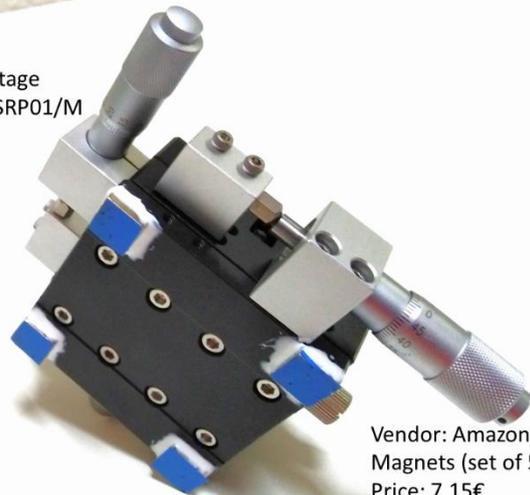

Vendor: Amazon
Magnets (set of 50)
Price: 7.15€

4 magnets are glued on the base of the manual stage (with epoxy glue, Nural 26) and a layer of insulator tape is attached.

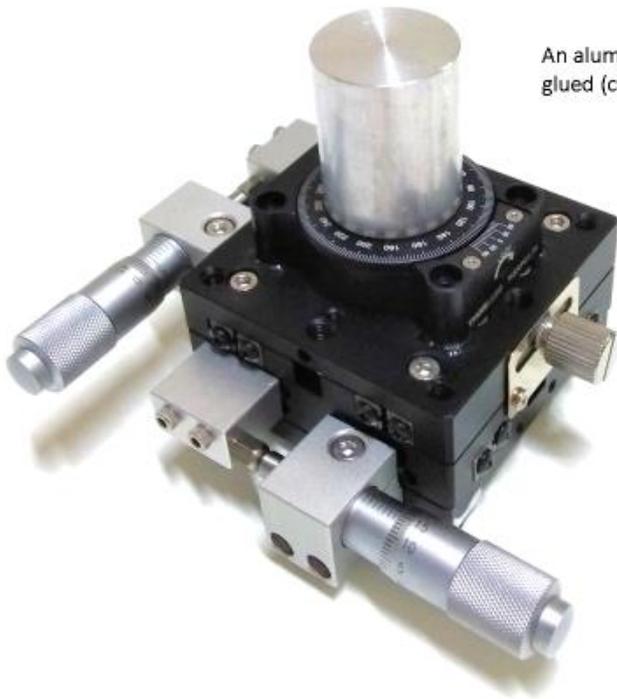

An aluminium machined cylinder post is glued (cyanoacrylate) onto the rotation stage.

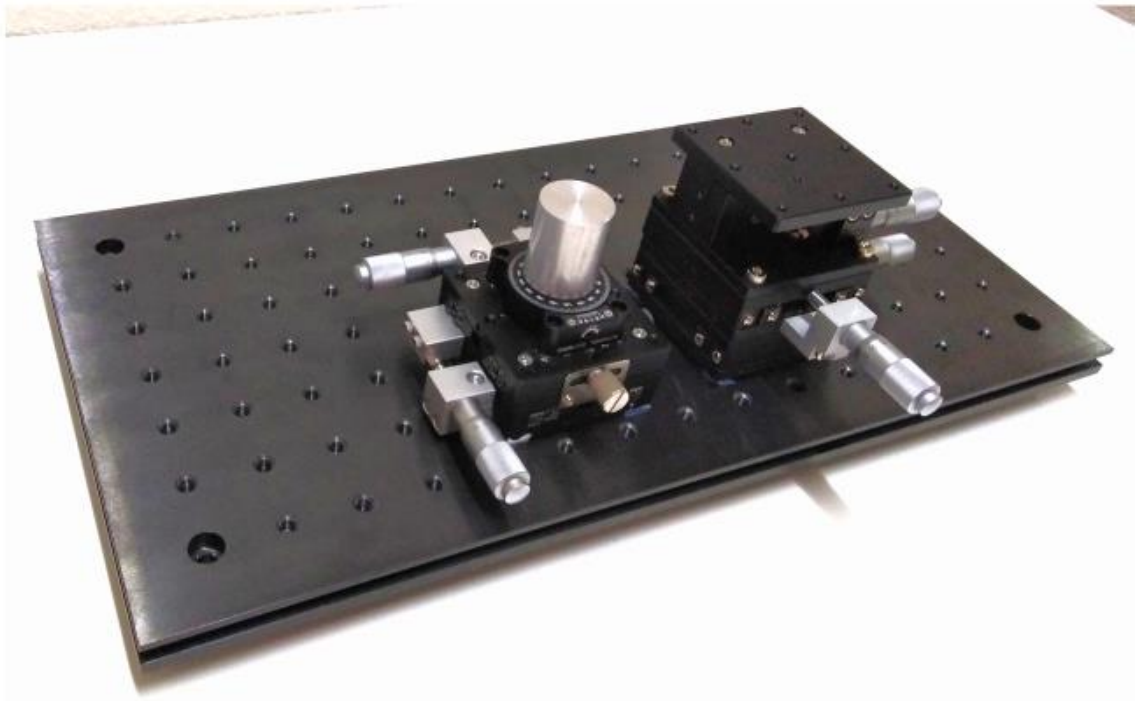

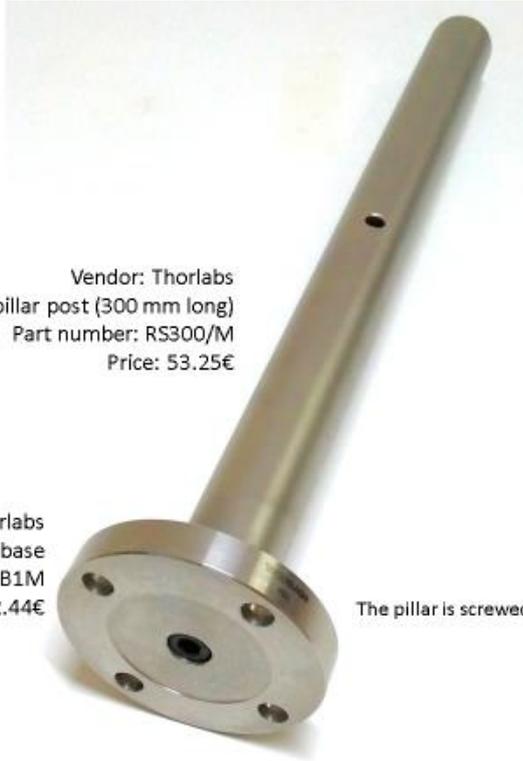

Vendor: Thorlabs
Ø25.0 mm diam pillar post (300 mm long)
Part number: RS300/M
Price: 53.25€

Vendor: Thorlabs
Mounting post base
Part number: PB1M
Price: 22.44€

The pillar is screwed to the base with a M6 screw

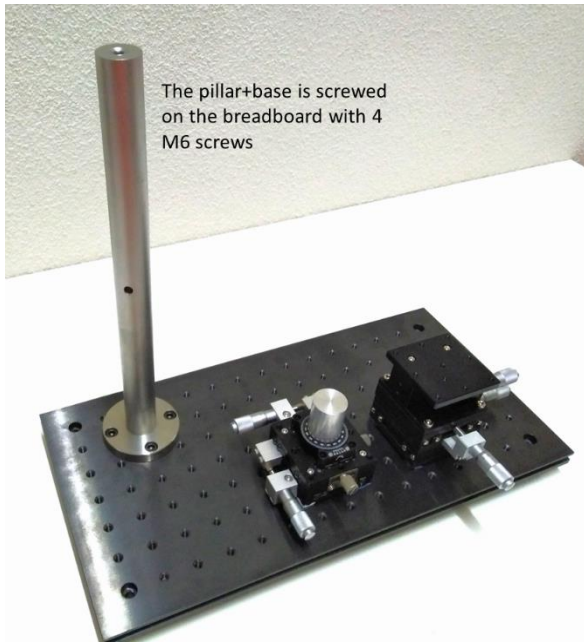

The pillar+base is screwed on the breadboard with 4 M6 screws

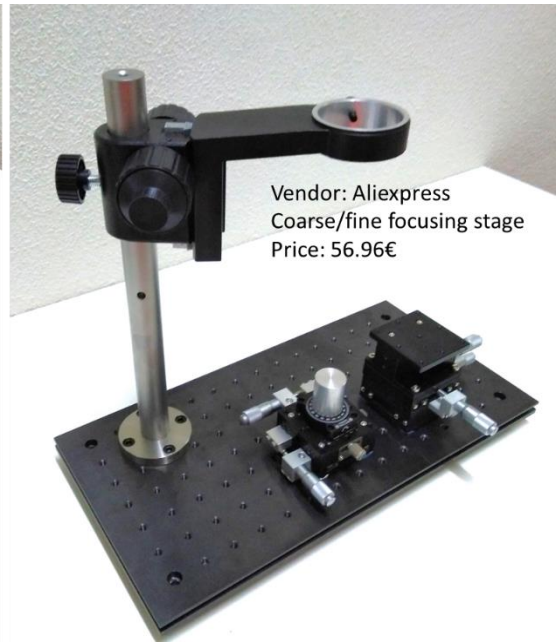

Vendor: Aliexpress
Coarse/fine focusing stage
Price: 56.96€

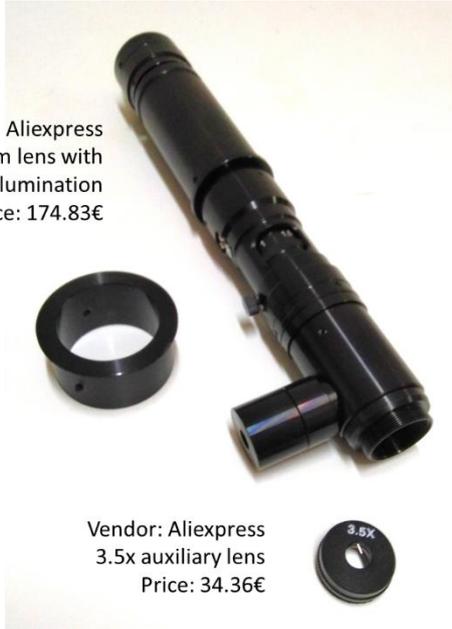

Vendor: Aliexpress
400x zoom lens with coaxial illumination
Price: 174.83€

Vendor: Aliexpress
3.5x auxiliary lens
Price: 34.36€

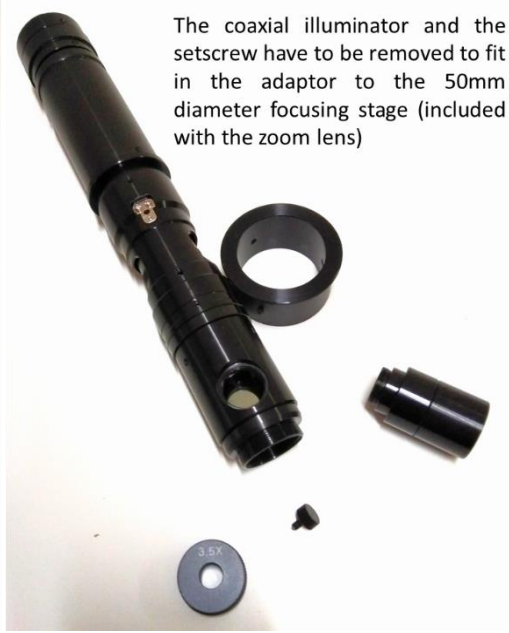

The coaxial illuminator and the setscrew have to be removed to fit in the adaptor to the 50mm diameter focusing stage (included with the zoom lens)

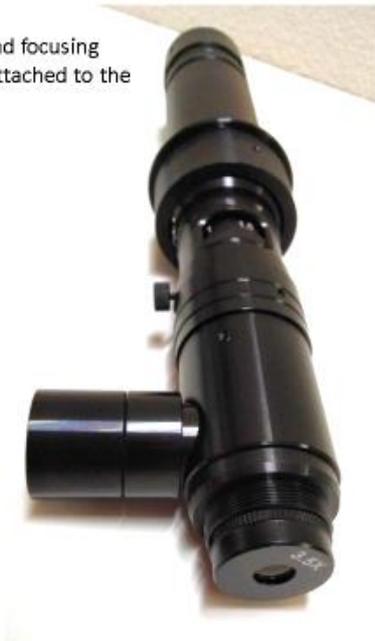

Auxiliary lens and focusing stage adaptor attached to the zoom lens.

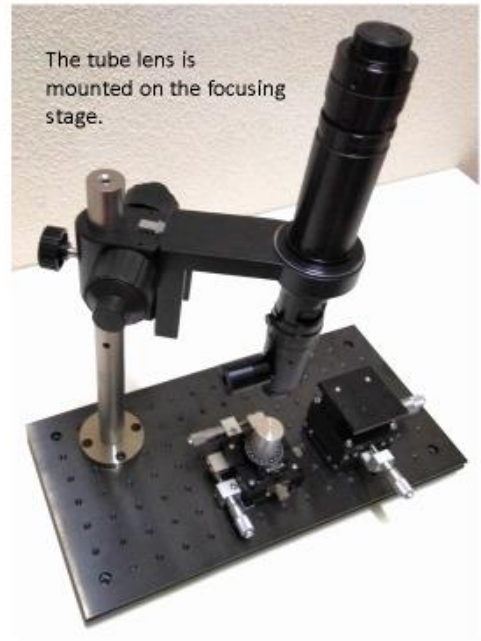

The tube lens is mounted on the focusing stage.

LED illuminator for the coaxial illumination
400x zoom lens (included with the zoom lens)

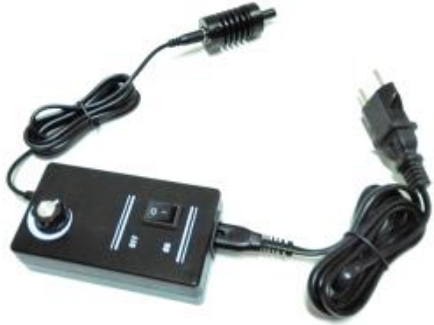

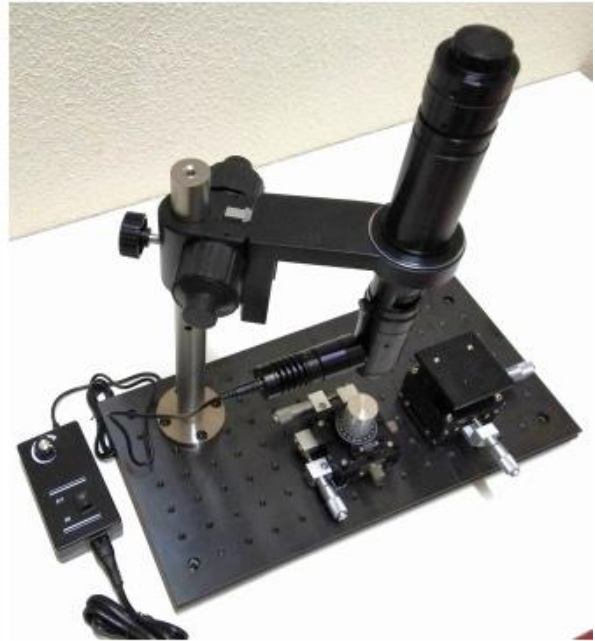

Vendor: Aliexpress
21 MPix digital camera USB + HDMI
Price: 74.70€

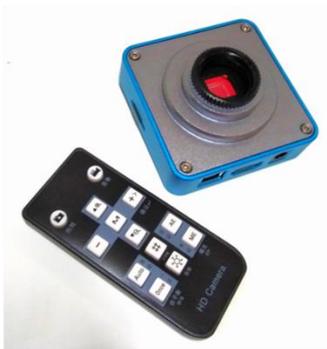

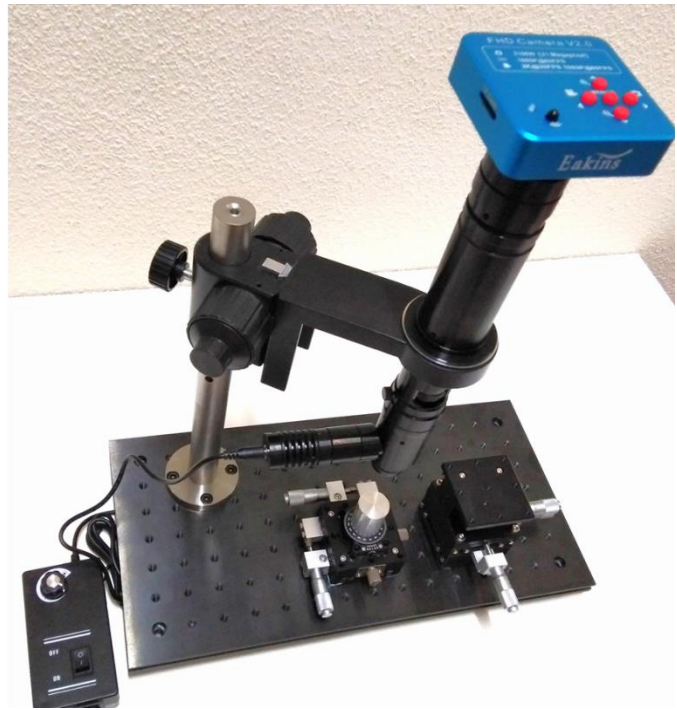